\documentclass{ckm}                 

\usepackage{txfonts}            

\confname{Workshop on the CKM Unitarity Triangle, IPPP Durham, April 2003}

\title{$B_s$ Physics and Prospects at the Tevatron}
\author{Donatella Lucchesi \address{University and INFN of Padova, Padova - Italy}\\
for the CDF and D0 Collaborations}
\begin{document}
\begin{abstract}
Both experiments CDF and D0 at the Tevatron collider have now the first samples of $B_s$ particles where preliminary measurements  are performed.   The mass and lifetime measurements are presented and  the yield  for the hadronic $B_s$ decays, the first step toward the $B_s$ production fraction and branching ratio measurements,  is discussed. This also sets the bases for a re-evaluation of mixing capabilities in Run II.
\end{abstract}
%
\maketitle
\section{Introduction}
The new data taking started March 2001,  marked a new era for the $B_s$ physics at the Tevatron collider. Both experiments, CDF and D0, have new strategies to select $B_s$ events among the huge $p\bar p$ production using triggers based  on the track impact parameter.  Fully hadronic decays like $B_s\rightarrow D_s\pi$, considered fundamental to measure the mixing frequency, can be selected requiring at least two tracks displaced from the primary vertex. \\
The combination of one displaced track and a lepton (muon or electron) allows to lower the threshold on the lepton trigger with respect to the Run I data taking, increasing the number of reconstructed events in the semileptonic sample.  But also the ``old'' dilepton trigger has been improved thanks to the new tracking systems and the upgraded  electronics increasing the number of events with a $J/\psi$ particle in the final state.  
Besides the mixing frequency many other measurements can be done in the $B_s$ system to test the Standard Model and to go beyond it. Here the mixing, the lifetime, the lifetime differences and the mass are discussed.
\section{$B_s$ mass, lifetime and lifetime difference}
\label{sec:life}
\begin{figure}
\hbox to\hsize{\hss
\includegraphics[width=\hsize]{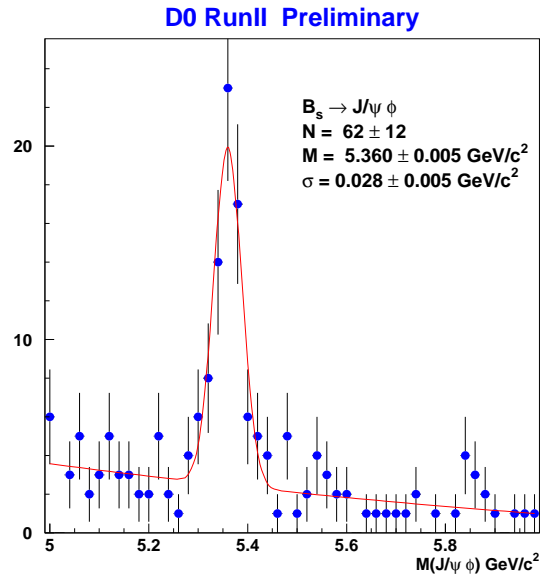}
\hss}
\caption{D0: invariant mass distribution of $J/\psi \Phi$ system.}
\label{fig:massD0}
\end{figure}

The precise $B_s$ mass determination is one of the measurements that both CDF and D0 can perform to
test  QCD. 
In this analysis the $B_s$ is reconstructed through the decay $B_s\rightarrow J/\psi \Phi$ with the $J/\psi\rightarrow \mu^+\mu^- $ and the $\Phi\rightarrow K^+K^-$.
Starting from the dataset of $\sim$ 47 pb$^{-1}$ collected with the dimuons trigger  D0 has reconstructed a sample of $62\pm12$ events, shown in Fig.~\ref{fig:massD0}, with a very low background.

\begin{figure}
\hbox to\hsize{\hss
\includegraphics[width=\hsize]{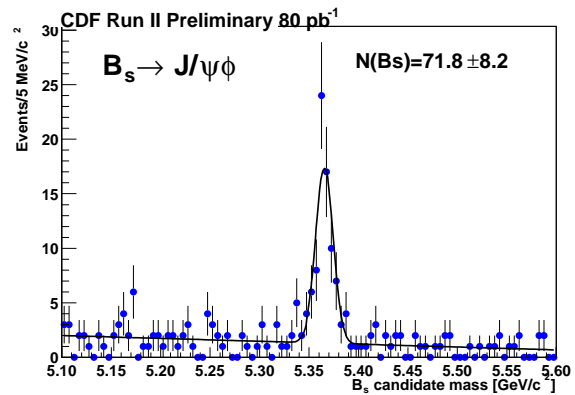}
\hss}
\caption{CDF: invariant mass distribution of $J/\psi \Phi$ system.}
\label{fig:massCDF}
\end{figure}

CDF reconstructs the same decay in a sample of  $\sim 80$ pb$^{-1}$. The  $71\pm8$ events (see  Fig.~\ref{fig:massCDF}) are used to measure the $B_s$ mass:  M($B_s$) = 5365.50$\pm$1.29(stat.)$\pm$0.94(syst.) MeV/c$^2$. 
This measurement  improves  the Run I~\cite{Bsmasscdf} statistical and systematical errors and it is, at the moment, the world best measurement. 

\begin{figure}
\hbox to\hsize{\hss
\includegraphics[width=0.5\hsize]{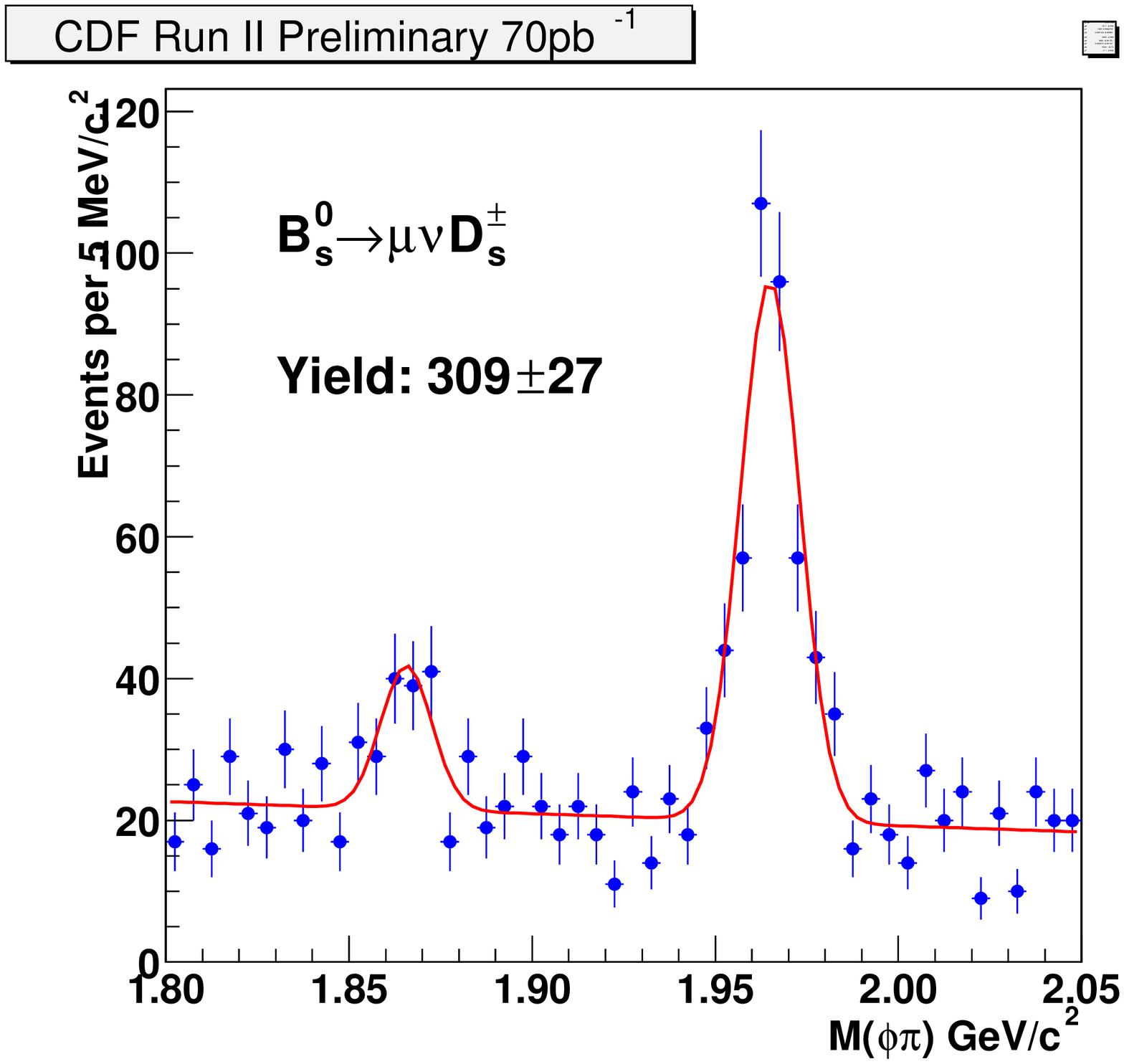}
\hss
\includegraphics[width=0.5\hsize]{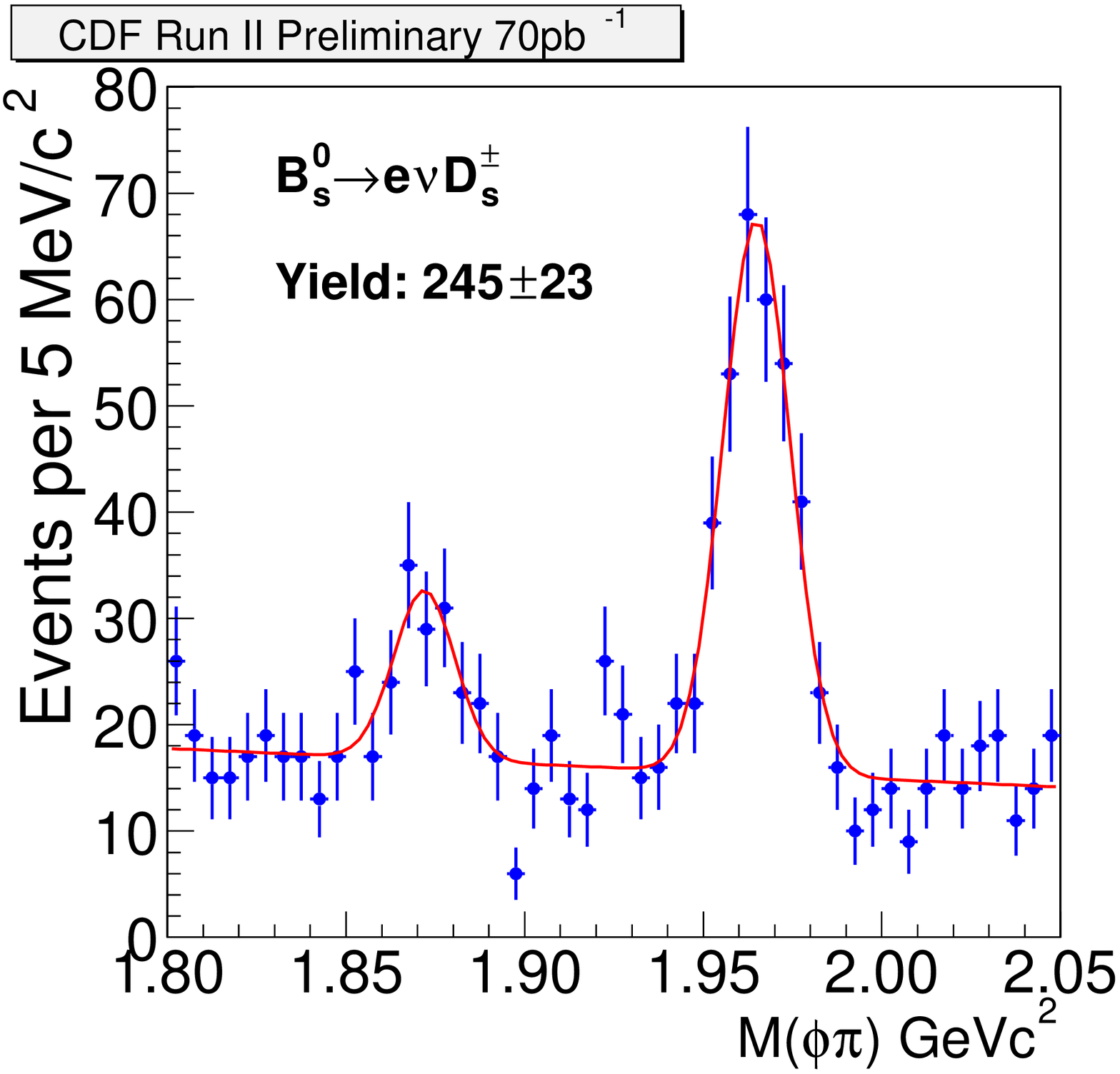}
\hss}
\caption{Invariant mass of $\Phi\pi$ when they are close to a lepton (muon and electron) and the charge combination is that required by the decays.}
\label{fig:lifeBs}
\end{figure}

The $B_s$ lifetime is expected~\cite{CKM1} of the same order as the $B_{d,u}$ lifetime and with the current data $B_s$ and $B_d$  lifetimes are consistent  within $\simeq$ 4\% accuracy. The $B_s$ measurements are dominated by statistical errors  and the Tevatron experiments are expected to improve them in a near future.\\ 
CDF can perform this measurement using two different samples: the semileptonic decays and the fully reconstructed hadronic decays.   The former has already been used at CDF in Run I~\cite{Bslife} and based on this analysis CDF  predicts that in 2 fb$^{-1}$ CDF can reach a 2$\%$ precision on the lifetimes. Fig. ~\ref{fig:lifeBs} displays the invariant mass when a $D_s\rightarrow \Phi\pi$ is reconstructed close to  a muon(left) and an electron(right) and the  charges of $D_s$ and lepton  match the requirement of the decay. The lifetime measurement is in progress. At the moment CDF has already $\sim$ 5 times the statistics of Run I with the signal to noise  ratio even better than expected. This has been achieved thanks to the trigger on the impact parameter since these candidates are reconstructed on the lepton plus displaced track sample.

The Run II predictions for the lifetime measured with  fully hadronic decays are based on the Monte Carlo simulation as described in~\cite{YB}. The statistics is expected to be larger than the semileptonic sample and with the proper time resolution  of  about 45 fs an error of $0.5\%$ with 2fb$^{-1}$ of data is within the CDF reach.

The $B_s\rightarrow J/\psi \Phi$ has two components:  one with CP even (dominant) with a short lifetime and one at CP odd with a long lifetime. These events will be used   to measure the lifetime difference, $\Delta\Gamma$, between the two eigenstates, the even and the odd. The best accuracy will obtained when the short component  measured in this decay and the average lifetime obtained from the $B_s\rightarrow D_s\pi$ are combined. 
By using the result obtained in Run I~\cite{DG1} CDF evaluated its reach with 2fb$^{-1}$ of data: $\Delta\Gamma/\Gamma \sim 0.06$.\\
At the moment the statistics is limited and  CDF presents a  measurement of the lifetime  without  separating the two components: $\tau(B^0_s)=1.26\pm0.20(stat)\pm0.02(sys)$.

This process is considered a ``gold-plated''  $B_s$ decay because it is also sensitive to new physics beyond Standard Model. Within this model the  CP asymmetry is expected to be very small and an  observation of a deviation can only be attributed to sources of new physics.
\section{$B_s$ mixing with semileptonic decays}
The mixing frequency in the $B_s$ system is one of most important measurement to verify the Standard Model. The current limit is $\Delta m_s>14.4$ ps$^{-1}$ which translates in $x_s=\Delta m_s \tau_{B_s}>21$. Comparing  this number  to the $B_d$ oscillation frequency $\Delta m_d=0.502\pm0.006$ ps$^{-1}$ gives an idea of how difficult will be to measure these fast oscillations.\\
The sensitivity on $x_s$ can be expressed: 
$$
sig(x_s) = \sqrt{\frac{N\epsilon D^2}{2} } exp(-\left(x_s\sigma_{c\tau}/\tau\right)^2/2) \sqrt{\frac S{S+1}}
$$
where N is the total number of events before the flavor tagging, $\epsilon$ is the tagging efficiency, $D$ the dilution, $\sigma_{c\tau}$ the proper time resolution and $S$ the signal-to-noise ratio.
The above formula drives the analysis: a sample of events with a reasonable signal-to-noise is needed, but also a good resolution in reconstructing the decay time is necessary as well as an efficient flavor tagging algorithm.
 
CDF set a limit $\Delta m_s>5.8$ ps$^{-1}$ at 95$\%$ C.L. using Run I $B_s$ semileptonic decays~\cite{mix1}. In Run II the statistics and the signal to noise ratio will be improved as already discussed in~\ref{sec:life} but the CDF sensitivity for $x_s$ in this decay channel remains limited due to the fact that $B_s$ is partially reconstructed .  The proper time $ct= \frac{L_{xy}M(B_s)}{P_t(B_s)} = {{L_{xy}M(B_s)}\over{P_t(l+D_s)}}\cdot K$ is given by the decay length divided by $\beta\gamma$, where $K$ is the correction factor for  the missing neutrino. This correction, calculated using the Monte Carlo, introduces an additional resolution factor, $\sigma_t = 60$fs $\oplus t \cdot \sigma_K/K$ with $\sigma_K/K\sim 14\%$. 
\begin{figure}
\hbox to\hsize{\hss
\includegraphics[width=0.8\hsize]{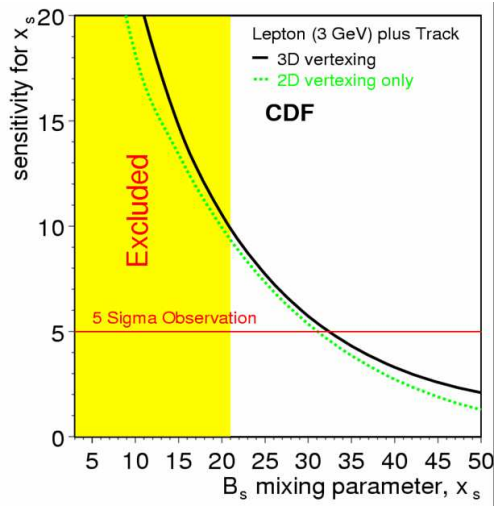}
\hss}
\caption{Sensitivity for $x_s$ using only $B_s$ semileptonic decays.}
\label{fig:mixing_semi}
\end{figure}
The CDF predictions, shown in Fig. ~\ref{fig:mixing_semi}, are calculated assuming the above time resolution in a sample of $\sim 40,000$ events (in 2 fb$^{-1}$) including $B_s\rightarrow D_s l$ and $B_s\rightarrow D_s^* l$, with $D_s\rightarrow \Phi \pi, K^{*0}K^{\pm}$ and $\pi^-\pi^+\pi^{\pm}$. The tagging figure of merit, $\epsilon D^2$ is assumed $11.3\%$. The conclusion is that this sample can be used to measure $x_s$ up to $\sim 30$.

The same exercise has been done by D0. The number of events assumed in 2 fb$^{-1}$ is 40,000 collected by triggering on two leptons, one from the reconstructed $B_s$ and the other from the other $B$ of the event.  D0 reach, $x_s\leq 30$, is similar to the CDF one.
\section{$B_s$ mixing with hadronic decays}
\label{sec:hadro}
Both experiments CDF and D0 in this new data taking period optimized their triggers to collect samples of hadronic $B_s$ decays where the it is fully reconstructed  in particular for the $B_s$ oscillations observation.

\begin{figure}
\hbox to\hsize{\hss
\includegraphics[width=0.5\hsize]{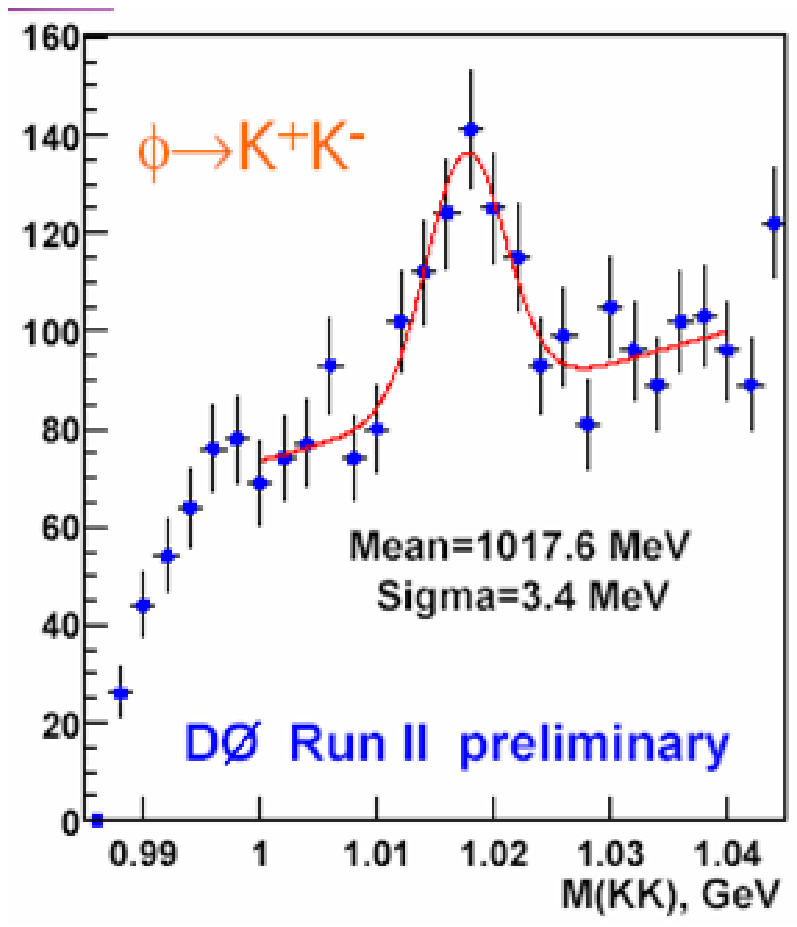}
\includegraphics[width=0.5\hsize]{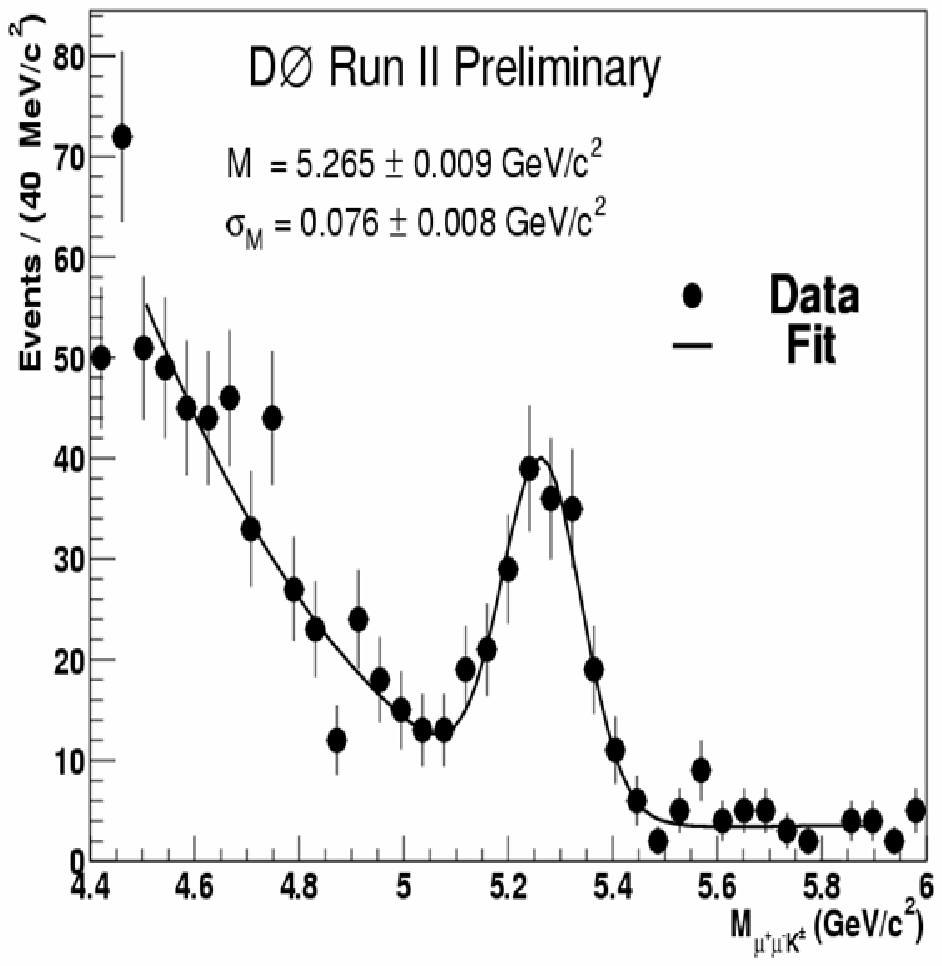}
\hss}
\caption{Left: D0 invariant mass of two kaons. Right: D0 invariant mass of $J/\psi$ K,  this sample is used to measure the tagging purity and efficiency.}
\label{fig:phiD0}
\end{figure}

D0 is planning to collect hadronic $B_s$ particles, triggering on the lepton from the other $B$ hadron in the events and by using a track trigger not implemented yet. 
D0 expects to reconstruct $\sim 1,000$ of $B_s\rightarrow D_s^{\mp}\pi^{\pm}$ and $B_s\rightarrow D_s^{\mp}\pi^-\pi^+\pi^{\pm}$ with $D_s^{\mp}\rightarrow\Phi\pi^{\mp}, K^{*0}K^{\mp}, K^{*\mp}K^0$. A first encouraging $\Phi$ resonance has been reconstructed, as shown in Fig. ~\ref{fig:phiD0}(left) in 40 pb$^{-1}$ of data.
The flavor tagging, which includes different algorithms, has been optimized on Monte Carlo. The purity and the efficiency are measured in a sample of $B^{\pm}\rightarrow J/\psi K^{\pm}$(see Fig. ~\ref{fig:phiD0} right) and compared with the predictions~\cite{YB}. 
\begin{table}[h]
\begin{center}
 \begin{tabular}{|l|c|c|}  \hline  
                              & Muons            & Jet Charge \\ \hline 
$\epsilon (\%)$ & $8.2\pm 2.2$  & $55.1 \pm 4.1$ \\
D$(\%)$            & $63.9\pm30.1$&$21.1\pm10.6$ \\
$\epsilon D^2$ & $3.3\pm1.8$ ($3.1$ pred. )   & $2.4\pm1.7$($4.7$ pred.) \\ \hline
\end{tabular}     
\caption{Summary of flavor tagging study at D0.}
\label{tab:tag}
\end{center}
\end{table} 
Table~\ref{tab:tag} summarizes the preliminary results regarding the efficiency, the dilution and the figure of merit, $\epsilon D^2$.

Exploiting its good  dilepton trigger D0 can collect between 400 and 1,000 events of $B_s\rightarrow J/\psi K^*$, $K^{*}\rightarrow K^{\pm}\pi^{\mp}$ which can be combined with the hadronic sample to increase the statistics.
\begin{figure}
\hbox to\hsize{\hss
\includegraphics[width=\hsize]{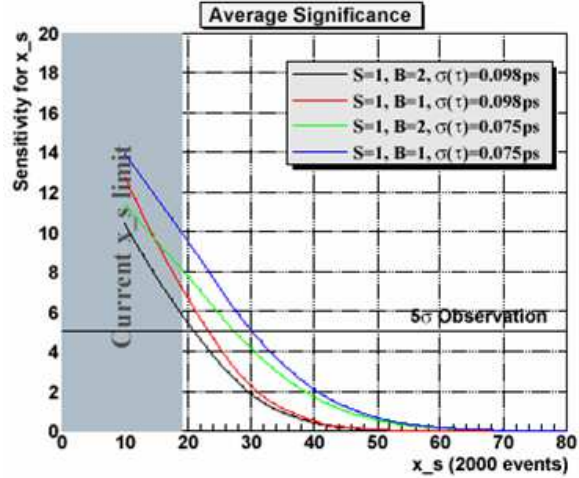}
\hss}
\caption{D0: $x_s$ sensitivity in fully reconstructed $B_s$ decays.}
\label{fig:D0mix}
\end{figure}

Under the assumption that the proper time resolution is $0.40$ ps, the estimated  sensitivity curve as function of $x_s$, is shown in Fig. ~\ref{fig:D0mix}. D0 can  measure $x_s$ up to 30.

\begin{figure}
\hbox to\hsize{\hss
\includegraphics[width=\hsize]{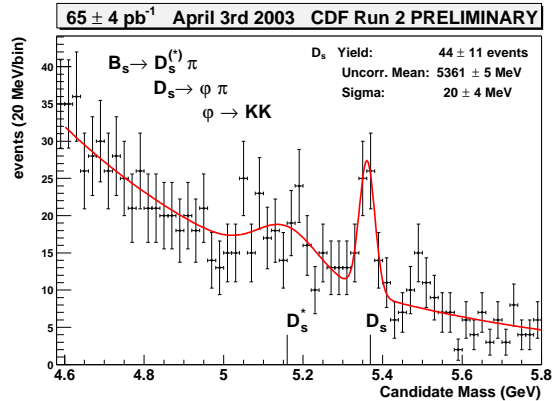}
\hss}
\caption{CDF: invariant mass of $D_s\pi$ reconstructed in the hadronic triggered sample.}
\label{fig:Bs}
\end{figure}

CDF has reconstructed for the first time the decay $B_s\rightarrow D_s^{\mp}\pi^{\pm}, D_s^{\mp}\rightarrow \Phi\pi^{\mp}$. In 65 pb$^{-1}$ of data collected triggering on displaced tracks $44\pm 11$ events are identified, see Fig. ~\ref{fig:Bs} where the $D_s\pi$ invariant mass is shown.
This is the first step to check the prediction on the $x_s$ sensitivity. While other parameters like the proper time resolution and the flavor tagging can be tested on other samples, the yield, namely the $B_s$ cross section times the branching fraction, is unknown and it is the first thing to measure.
The direct measurement of $N_{B_s}$ goes through the knowledge of the $b$ cross section and the $B_s$ fraction, $N _{B_s}= \sigma(b)f_sBr(B_s\rightarrow D_s^{\mp}\pi^{\pm})Br(D_s^{\mp}\rightarrow \Phi\pi^{\mp})Br(\Phi\rightarrow K^+K^-)$. Both quantities, $\sigma(b)$ and $f_s$ are not well known. The $b$ production cross section has a 2 sigma discrepancy between data and theory while for $f_s$  the LEP determination differs from the Tevatron measurement~\cite{PDG,fs} and in principle can be different. These two facts motivated CDF to measure a relative $B_s$ yield. The best decay channel to normalize to is $B_d\rightarrow D^{\mp}\pi^{\pm}$ with $D^{\mp}\rightarrow K^{\mp}\pi^{\pm}\pi^{\pm}$ also reconstructed in the hadronic triggered data. By using these two decays CDF can measure:
\begin{eqnarray*}
\frac{N(B_s)}{N(B_d)} = &\frac{f_s}{f_d} \frac{\epsilon(B_s)}{\epsilon(B_d)}\times \hspace{3truecm} \\ 
  &  \frac{Br(B_s\rightarrow D_s^{\mp}\pi^{\pm})Br(D_s^{\mp}\rightarrow \Phi\pi^{\mp})Br(\Phi\rightarrow K^+K^-)}{Br(B_d\rightarrow D^{\mp}\pi^{\pm})Br(D^{\mp}\rightarrow K^{\mp}\pi^{\pm}\pi^{\pm})} 
\end{eqnarray*}

\begin{figure}
\hbox to\hsize{\hss
\includegraphics[width=\hsize]{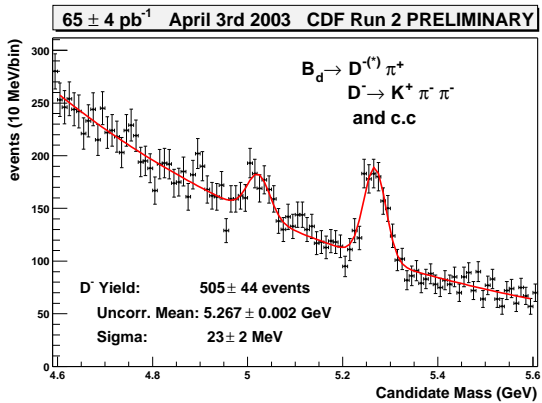}
\hss}
\caption{CDF: invariant mass of $D^{\mp}\pi^{\pm}$ as reconstructed in the hadronic triggered dataset.}
\label{fig:Bd}
\end{figure}

The number of $B_d\rightarrow D^{\mp}\pi^{\pm}$ identified is $N(B_d) = 505\pm 44$. The invariant mass of the $D^{\mp}\pi^{\pm}$ is displayed in Fig. ~\ref{fig:Bd}. The second peak around 5 GeV/c$^2$ is due to partially reconstructed $B_d\rightarrow D^{*\mp}\pi^{\pm}$ where a $\pi^0$ from  $D^{*\mp}\rightarrow D^{\mp}\pi^0$ is not reconstructed.

From the numbers of reconstructed events: $\frac{N(B_s)}{N(B_d)}= 0.087\pm 0.023$. A systematic error of $0.008\oplus0.008$ due to the fitting procedure has been assigned to the ratio for $B_s$ and $B_d$.

\begin{figure}
\hbox to\hsize{\hss
\includegraphics[width=\hsize]{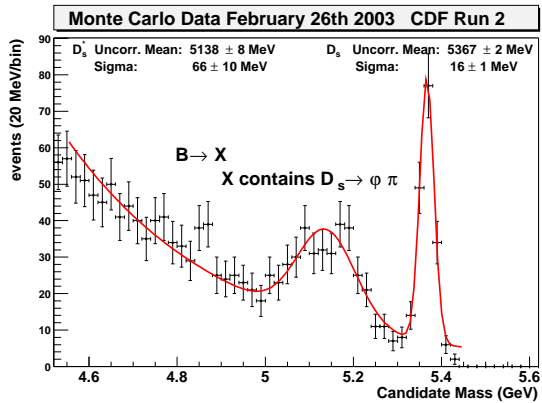}
\hss}
\caption{CDF: invariant mass of $D_s\pi$ reconstructed in Monte Carlo data after the trigger and detector simulation.}
\label{fig:MC}
\end{figure}

The relative efficiencies, $ \frac{\epsilon(B_s)}{\epsilon(B_d)}$ have been measured using Monte Carlo data. This ratio includes the relative trigger and  reconstruction efficiencies. A very detailed Monte Carlo detector simulation has been developed and tuned on data to reproduce the different data taking configurations CDF had: the silicon coverage now is almost 100$\%$ but in two years varied a lot due to Tevatron incidents and subsequent recovering. This is one of the most important elements to take into account to properly evaluate the efficiencies. On the other hand the trigger on displaced tracks depends on the silicon coverage and its optimization in reconstructing tracks has changed to follow the detector behavior. 
In Fig. ~\ref{fig:MC} the invariant mass of $D_s\pi$ on Monte Carlo data is shown after the trigger and the detector simulation have been performed demonstrating a good agreement with data. 

By using this procedure   $ \frac{\epsilon(B_s)}{\epsilon(B_d)}= 1.08\pm 0.02(stat)^{+0.06}_{-0.08}$ is obtained. \\
The major sources of systematic errors on this ratio are the $B_s$, $B_d$, $D_s$ and $D^{\mp}$ lifetimes and the $b$ quark momentum spectrum.

The ratio $\frac{Br(D_s^{\mp}\rightarrow \Phi\pi^{\mp})Br(\Phi\rightarrow K^+K^-)}{Br(D^{\mp}\rightarrow K^{\mp}\pi^{\mp}\pi^{\pm})}= 0.19\pm 0.05$ is taken from PDG~\cite{PDG}.
By combining all the numbers CDF quotes:
$$
\frac{f_s}{f_d}\frac{Br(B_s\rightarrow D_s^{\mp}\pi^{\pm})}{Br(B_d\rightarrow D^{\mp}\pi^{\pm})} = 0.42 \pm 0.11 (stat) \pm 0.11(PDG) \pm 0.07(syst)
$$

This is the first measurement of the $B_s$ branching ratio times the production fraction,  which is dominated at present by the statistical error and the uncertainties on the $D$ branching ratios.
\section{Conclusions}
CDF and D0 have reconstructed the first signals of $B_s$ decays in data collected in the new data taking period. Both collaborations are working in order to optimize the tools necessary to perform the various measurements accessible in this sector, in particular a big effort is going in the mass and  lifetime determination and of course in the mixing frequency analysis.
Both D0 and CDF  have  nice samples of $B_s\rightarrow J/\psi\Phi$ where the mass has been measured by CDF.
While the lifetime determination is in progress, CDF has shown for the first time the decay $B_s\rightarrow D_s\pi$ where the relative production fraction has been measured. 
Both experiments have to face the fact the the Tevatron is giving a lower luminosity than  expected and that the sensitivity quoted in~\cite{YB} result to be  somehow optimistic. The first thing is not so negative as can appear for $B$ physics because the trigger threshold optimized for high rate can be lowered. This is currently under study in CDF, in particular for the trigger on displaced tracks. 
The other good thing is that other $B_s$ decays were not considered  in ~\cite{YB} and thus the $B_s$ statistics can be increased.
Work is in progress to quantify all these factors and give an updated expectation  for a measurement of $B_s$ mixing.


\end{document}